\documentclass[aps,prl,preprint,groupedaddress]{revtex4}

\usepackage{epsfig}
\usepackage{amssymb}
\usepackage{times}
\usepackage{amsmath}
\usepackage{color}

\begin{document}

\def\lambar{\lambda \hspace*{-5pt}{\rule [5pt]{4pt}{0.3pt}} \hspace*{1pt}}

{\Large  DESY-24-091}

{\Large  June 2024}

\bigskip
\bigskip
\bigskip
\bigskip


\title{Two-bunch seeding of soft X-ray free electron lasers}



\author{E.~Schneidmiller}
\email{evgeny.schneidmiller@desy.de}
\author{I.~Zagorodnov}
\email{igor.zagorodnov@desy.de}
\affiliation{\small Deutsches Elektronen-Synchrotron DESY, Notkestr. 85, 22607 Hamburg, Germany}

\date{\today}

\begin{abstract}
Seeded Free Electron Lasers (FELs) demonstrate a good performance and are successfully used in different user experiments in extreme ultraviolet and soft X-ray regimes. In this paper a simple modification of the seeding 
scenario is proposed relying on generation of two closely spaced bunches with very different properties: a low-current seeding bunch, and a high-current bunch that amplifies coherent radiation, produced by the seeding bunch. 
This approach eliminates different limitations and mitigates some harmful effects in the standard scenario. In particular, one can generate very high harmonic numbers with a moderate laser power in a simple high-gain harmonic generation (HGHG) scheme. Alternatively, in case of moderate harmonic numbers, one can strongly reduce the required laser power thus simplifying design of high repetition rate seeded FELs. 
An influence of beam dynamics effects (like nonlinearities of the longitudinal phase space of electron beams, coherent synchrotron radiation, longitudinal space charge, geometrical wakefields, microbunching instabilities etc.) on properties of output radiation (spectrum broadening, pedestals, stability) can be to a large extent reduced in the proposed scheme. In this paper we illustrate the operation of the two-bunch seeding scheme in HGHG configuration with realistic start-to-end simulations for the soft X-ray user facility FLASH. We show that nearly Fourier-limited multi-gigawatt pulses can be generated at 4 nm using the present compact design of the undulator system. With several thousand pulses per second this can be a unique source for photon science. 
\end{abstract}


\maketitle

\section{Introduction}

Implementation of seeding schemes in high-gain short-wavelength FELs promises exceptionally bright output radiation with stable properties \cite{fermi}. There are two main schemes that have been proposed and tested: High-Gain Harmonic Generation (HGHG) \cite{yu} and Echo-Enabled Harmonic Generation (EEHG) \cite{stupakov}. 

In case of HGHG, an electron beam is modulated in energy by a laser in a modulator undulator followed by a chicane where the energy modulations are converted into density modulations. The latter are nonlinear, i.e. the electron density contains higher harmonics that produce radiation in the following undulator (FEL amplifier) tuned to one of those harmonics. If this undulator is sufficiently long, the radiation is amplified up to FEL saturation. This scheme is conceptually and technically simple, it was successfully tested at different facilities and used in routine operation for users \cite{fermi}. However, there is an intrinsic limitation that does not allow to use this concept at very high harmonic numbers. In order to generate density harmonic number $n$ in the modulator-chicane system, one should impose the energy modulation $\Delta \cal{E}$ that is significantly larger than uncorrelated energy spread in the electron beam $\sigma_{\cal{E}}$ \cite{yu}:

\begin{equation}
\frac{\Delta {\cal E}}{\sigma_{\cal{E}}} > n  \ .
\label{hghg-condition}
\end{equation}  

One problem with this condition is that strong energy modulations at large harmonic numbers can prevent lasing in the amplifier. A possible method to deal with this issue is the so-called ``fresh bunch technique'' proposed in \cite{fresh-bunch}. Another problem is that the required laser power (for a given uncorrelated energy spread) scales as the square of harmonic number, 
$P_{L} \propto (\Delta {\cal E})^2 \propto n^2$,
what often makes generation of very high harmonics impractical. A possible approach to reach high harmonic number could be a modified HGHG setup that incorporates the optical klystron effect \cite{self-mod,ok-hghg} with a subsequent use of the fresh bunch technique \cite{self-enhanced}. However, this method requires exceptionally bright electron beams to reach short wavelengths. Moreover, an additional chicane, required for this scheme, should have a large longitudinal dispersion (characterized by the momentum compaction factor $R_{56}$) which can contribute to a development of the microbunching instability \cite{mbi-1,mbi-2,mbi-3}. 

The key parameter for HGHG scheme, as can be seen from Eq.~(\ref{hghg-condition}), is the uncorrelated energy spread. It is usually small in injectors of FEL drivers but then it increases proportionally to a peak current in bunch compression systems. High peak current is needed for a successful operation of the FEL amplifier at short wavelengths. This is an intrinsic contradiction of the present operating regimes of the seeded FELs: the same bunch with a high current and a relatively large uncorrelated energy spread is used to generate harmonics in modulator-chicane system, and then to produce the radiation at a harmonic in the amplifier. We propose to separate these functions by generation of two bunches in the accelerator that drives the FEL: a weakly compressed bunch with low current and low energy spread to be used in the modulator-chicane system, and a strongly compressed bunch to amplify the radiation produced by a weakly compressed bunch. The main goal of our studies is to understand if such bunches can be simultaneously produced in a typical accelerator used as FEL driver. We use parameters of accelerator of the first soft X-ray FEL user facility FLASH \cite{flash-nat-phot, njp} being upgraded towards high repetition rate seeded FEL facility \cite{fl2020plus}. 

Apart from solving the main conceptual problem of seeded FELs, as briefly discussed above, two-bunch approach has additional advantages: improved spectral brightness and stability of FEL radiation, relaxed requirements to laser power (which can be important for high repetition rate FELs) etc. In this paper we concentrate on HGHG scheme and only briefly discuss EEHG. The EEHG concept is more sophisticated than HGHG, it involves two lasers, two modulators and two chicanes \cite{stupakov}. Although the condition (\ref{hghg-condition}) is not applied to EEHG case, the required energy modulations by lasers are still proportional to uncorrelated energy spread, i.e. EEHG method can also strongly profit from two-bunch concept.  

We should note that two-bunch self-seeding of X-ray FELs \cite{mev} was proposed in \cite{two-bunch-ss}. In that scheme the bunches were separated by one RF cycle and had identical properties. For external seeding, considered in this paper, we need to generate two high-quality bunches with very different properties, and a separation must be smaller than one RF wavelength.  This is not a trivial task but we performed comprehensive start-to-end simulations of FLASH demonstrating that a solution to this problem can be found in a given accelerator system.

\section{The concept of two-bunch seeding}

\begin{figure}[tb]

\includegraphics[width=1.0\textwidth]{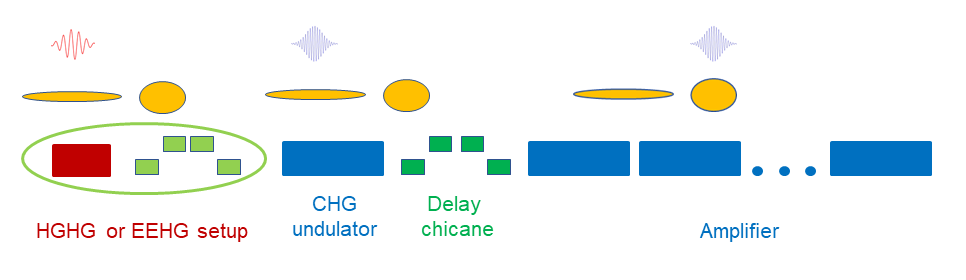}

\caption{
Conceptual representation of two-bunch seeding scheme.
}
\label{fig:2b-scheme}
\end{figure}

Conceptual scheme of two-bunch seeding of a short-wavelength FEL is shown in Fig.~\ref{fig:2b-scheme}.
The scheme looks similar to that proposed in \cite{fresh-bunch} with the difference that we want to use two bunches with different properties.
Therefore, we start the discussion with the generation of two bunches in an accelerator system. A typical driver of a high-gain short-wavelength FEL consists of a laser-driven RF gun, RF accelerating sections, bunch compressors and, eventually, a laser heater. The only hardware modification required for operation of two-bunch scheme is the installation in a laser system of a split-and-delay unit with an attenuator in one of the branches. Alternatively, two lasers can be used as in the case of FLASH. Two bunches with, generally speaking, different charges and controllable time delay are extracted from the cathode of the RF gun and are then compressed in a different way since they propagate in different fields of RF cavities. They arrive at the entrance of seeded FEL such that the low-current bunch (which we will call Seeding bunch or S-bunch) is behind the high-current bunch (Amplifying bunch or A-bunch). The S-bunch with low uncorrelated energy spread is modulated by a seed laser with a subsequent conversion of energy modulation into density modulation (HGHG scheme) or is  
being manipulated in a more complex way with the help of two lasers (EEHG). The S-bunch, containing higher harmonics of the laser in density distribution, then produces coherent radiation at a harmonic in a relatively short undulator (this is often referred to as Coherent Harmonic Generation, or CHG). In the following delay chicane, both bunches are delayed with respect to the radiation pulse which is parked on the A-bunch and is amplified to saturation in a long amplifier. The bunching in the S-bunch is smeared in the chicane, so that this bunch can only produce SASE (Self-Amplified Spontaneous Emission \cite{book}) in the amplifier but due to its low current the gain is low, and the generated background is negligible. Below in this Section we will briefly discuss different aspects of the two-bunch scheme concerning the formation of bunches in the accelerator and FEL operation.

\subsection{Beam dynamics}

\subsubsection{The photoinjector}

The photoinjector laser system can be simply upgraded by adding a split-and-delay unit, for example of interferometer type (with adjustable delay and attenuator). Thus, two pulses with the same length but different intensities and a controllable delay are generated. Some FEL facilities like FLASH \cite{flash-nat-phot} and European XFEL \cite{winni} are equipped with two photoinjector lasers allowing to produce two pulses with different properties and controllable delay. The cathode of the RF gun is illuminated by two pulses, and two electron bunches with equal or different charges are produced. The RF phases of two pulses can differ significantly what potentially might be a challenge because emittances and Twiss parameters of these bunches can be essentially different. Transverse mismatch, generated in the injector and later in the linac, can be controlled and minimized independently for both bunches as explained below.

\subsubsection{Bunch compression}

Typical bunch compression system consists of one or two chicanes operated at different energies. To linearize compression process, a high harmonic RF cavity is typically used (3rd harmonic in case of FLASH). When optimizing compression of two bunches, we came to the conclusion that it is better to use overcompression regime, see the details below in this paper.
In particular, it means that the two bunches change their relative positions: S-bunch leaves the gun first but arrives at the FEL setup last. Since S-bunch is only weakly compressed, its properties are expected to be stable, and this will guarantee high spectral stability of the output FEL radiation. 

\subsubsection{Collective effects}

Operation of seeding schemes is sensitive to the collective effects in the beam formation and transport systems. Collective effects like space charge, coherent synchrotron radiation (CSR), geometrical wakefields in a linac lead to a degradation of beam quality, and the degradation is more significant for strongly compressed bunches.   
In our simulations, the most important collective effects are properly included but we do not see any essential deterioration of the longitudinal phase space of the S-bunch because it is weakly compressed. At the same time, the action of A-bunch on S-bunch is weak due to a significant separation of them in time. As it will be discussed in more detail below, the properties of the longitudinal phase space (LPS) of the S-bunch define the spectral quality of the output radiation so that it can be kept almost ideal.   

\subsubsection{Mismatch control}

One of the possible issues with the proposed scheme might be a strong difference in properties of the transverse phase spaces (characterized by Twiss parameters) of the two bunches. The reason is that they are formed and evolve under the action of very different transverse forces from RF fields as well as from collective fields (mainly space charge), especially in the injector.
The problem is solved by introducing two matching points in the machine. The first one is in front of the FEL setup where the S-bunch is matched. The second one, for the A-bunch, is between the CHG undulator and the amplifier where the matching can be done by quadrupoles before and after the delay chicane. Since bunches can be switched on and off independently, one can observe and analyze the image of the relevant bunch only. Fine matching can be done empirically while optimizing FEL performance.

\subsection{FEL operation}

\subsubsection{HGHG and EEHG}

There is a consensus among FEL physicists that seeding at high harmonic numbers should be done with EEHG scheme. Indeed, it does not have the limitation (\ref{hghg-condition}), so that for a given energy spread one can go to a much higher harmonic number than in the case of HGHG. At the same time, EEHG requires a chicane with a relatively large $R_{56}$ in the first stage which can create serious problems. The most significant issue is that appearance of this chicane results in one more amplification cascade for the microbunching  instability. Another problem is a distortion of the longitudinal phase space in this chicane due to Coherent Synchrotron Radiation (CSR) that can result in a reduced bunching factor \cite{samoilenko}. In addition, Intra-Beam Scattering (IBS) effects can deteriorate the process of generation of a large bunching factor at high harmonic numbers. The two-bunch seeding concept mitigates the mentioned issues due to a low current of the S-bunch. 

Moreover, the concept makes it possible to generate high harmonics in a relatively simple HGHG configuration. In this case the required $R_{56}$ is typically by two orders of magnitude smaller than in the case of the first stage of EEHG. Thus, such effects as additional microbunching instability gain or CSR in dipoles of the chicane do not play a significant role anymore. In this paper we concentrate on HGHG case leaving EEHG for future studies. Below we would like to qualitatively discuss potential advantages of the two-bunch seeding even if we do not aim at studying sensitivity and stability aspects in this paper. 

\subsubsection{Spectral quality and stability}

Spectral quality of seeding schemes is sensitive to the longitudinal phase space of electron bunches. In case of the standard HGHG scheme, the beam is modulated by a laser in the modulator undulator, and a nonlinear transformation of LPS, leading to the appearance of high harmonics of the laser in electron density, occurs in the chicane. Since LPS is never ideal, its imperfections are embedded in bunching at higher harmonics and lead to a degradation of spectral quality of the radiation produced in the amplifier. In particular, a linear energy chirp results in a shift of central wavelength. It is not critical by itself but may lead to shot-to-shot variation of the wavelength if the compression is not stable. Nonlinear chirp leads to a spectral broadening, moreover the jitter of the central wavelength can also occur if there is an arrival time jitter between the bunch and the laser pulse. High-frequency modulations of the electron bunch due to the microbunching instability can lead to the generation of sidebands, a pedestal etc. Note that all these effects are much weaker in the amplifier because its $R_{56}$ is much smaller than that of the chicane.

In the case of two-bunch seeding, a long low-current bunch is used for harmonic conversion in the modulator-chicane system, and the spectral quality and stability is much less affected. There is, in general, a linear chirp leading to the wavelength shift but this chirp is very stable due to a reduced compression factor. Nonlinearities of the LPS of S-bunch are relatively small, i.e. one should expect neither significant spectrum broadening nor the jitter of the central wavelength. Microbunching instability is also expected to play a less significant role in the case of low-current bunch.
High-current A-bunch can be stronger distorted due to collective affects and can be less stable due to a stronger bunch compression but, as it was just mentioned, the deteriorating effects in the amplifier are much weaker.

\subsubsection{Debunching in drifts and undulators}

The modulated electron beam after the modulator-chicane system can relatively quickly lose bunching at high harmonics due to nonlinear space charge forces in the drifts and, especially, in the undulators \cite{hemsing}. Recent experimental studies \cite{khan} in a standard HGHG setup show that the effect can lead to a significant debunching on a relatively short distances if the beam current is high. In case of two-bunch seeding the current of S-bunch is low, and the CHG undulator is not long, so that these effects are not expected to play any significant role. 

\subsubsection{High harmonic numbers or a reduced laser power}

Since the uncorrelated energy spread of the S-bunch is relatively small, according to the condition (\ref{hghg-condition}) we can either go for a high harmonic number or strongly reduce the required laser power in case of moderate harmonic numbers. If, for example, the energy spread in the S-bunch is reduced by an order of magnitude w.r.t. a standard case, the laser peak power can then be smaller by two orders.  
The laser power reduction can also be achieved with the optical klystron approach \cite{self-mod,ok-hghg} (even though the quality of the output radiation might have to be compromised due to an additional strong chicane). Here we would like to mention that both approaches, two-bunch seeding and optical klystron, can be combined thus leading to enormous reduction of the required laser power and making high repetition rate CW seeded FELs possible. 

In this paper we do not consider the laser power reduction, and concentrate on the generation of high harmonic numbers.

\section{Simulations of beam dynamics in FLASH accelerator}

FLASH \cite{flash-nat-phot} is the first free-electron laser for XUV and soft X-ray radiation. It covers a wavelength range from 4 nm to about 90 nm with GW peak power and pulse durations between a few fs and 200 fs. 
The electron bunches with maximum energy of 1.35 GeV are distributed between the two branches, FLASH1 and FLASH2 \cite{njp}. The facility is based on the superconducting accelerator which allows to operate in a ``burst mode'' with long pulse trains (several hundred pulses) at 10 Hz repetition rate.
Presently, the facility is being upgraded towards high repetition rate seeding in the FLASH1 branch \cite{fl2020plus}.

\begin{figure}[tb]
	\includegraphics[width=1.0\textwidth]{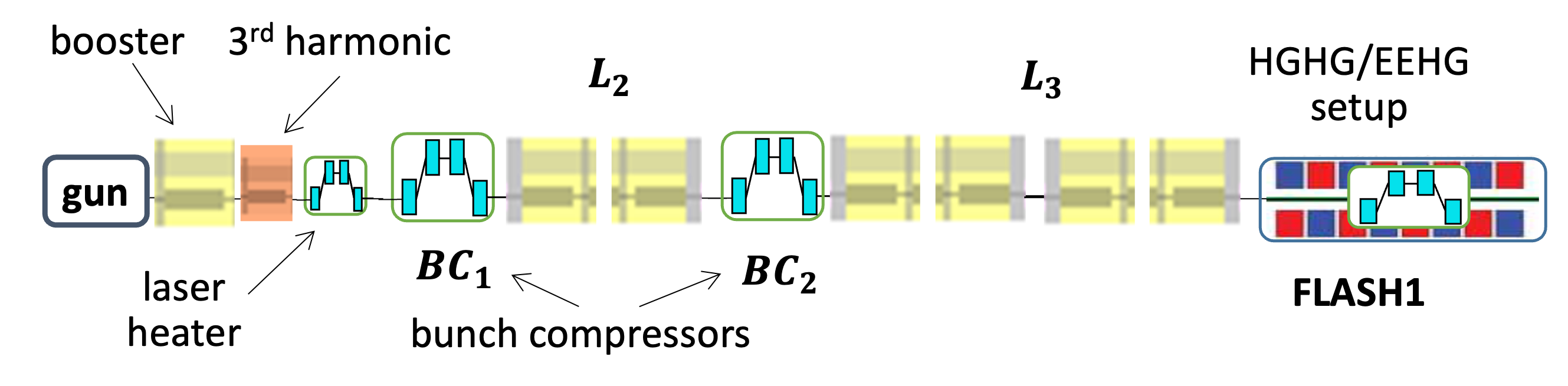}
	\caption{
		The layout of FLASH1.
	}
	\label{fig02}
\end{figure}

The layout of facility with FLASH1 branch is shown in Fig.~\ref{fig02}. In order to compress the beam to a high peak current the electron beam line incorporates two horizontal bunch compressors of C-type.

\begin{figure}[tb]
	\includegraphics[width=1.0\textwidth]{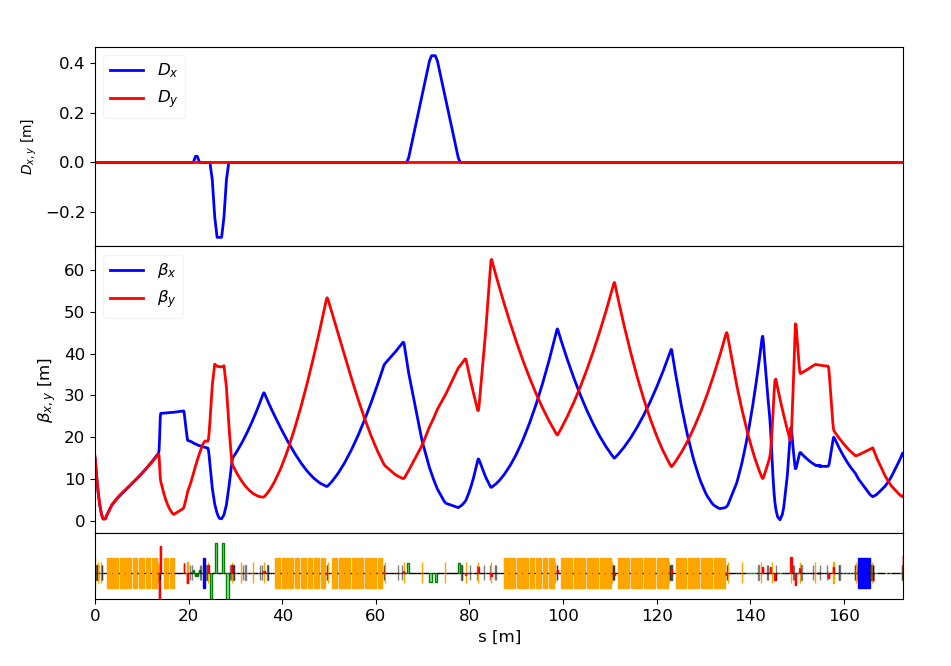}
	\caption{
		The design optics of FLASH1. The bottom plot presents an outline of different elements: quadrupoles (in red), dipoles (in green), RF modules (in orange). 
	}
	\label{fig03}
\end{figure}

The maximal accelerating voltages at different sections of the accelerator (as  it was assumed in the simulations) 
are listed in Table~\ref{TableRF} .  The ranges for the compression compaction factors are presented in Table~\ref{TableR56} .

\begin{table}[htbp]
	\centering
	\caption{Maximal accelerating voltage in RF sections.}
	\label{TableRF}
	\begin{tabular}{lcccll}
		\hline\hline
		{\bf section name}& 	{\bf booster}&	\boldmath{$3^{rd}$}{\bf harm.}& \boldmath{$L_2$}&	\boldmath{$L_3$}\\
		\hline
		{\bf maximal voltage, MV} &170&22.5&440&800\\
		\hline\hline
	\end{tabular}
\end{table}

\begin{table}[htbp]
	\centering
	\caption{Range of momentum compaction factors in the bunch compressors.}
	\label{TableR56}
	\begin{tabular}{lcccll}
		\hline\hline
		{\bf section name}& 	{\boldmath{$BC_1$}}&{\boldmath{$BC_2$}}\\
		\hline
		\boldmath $|R_{56}|$, {\bf mm} &120 - 250& 0 - 105 \\
		\hline\hline
	\end{tabular}
\end{table}

The linear optics is shown in Fig.~\ref{fig03}. It starts after the RF gun at the position $z=2.6$ m from the gun cathode. The position $z=2.6$ m corresponds to the beginning of the booster RF module with eight TESLA superconducting cavities (see Fig.~\ref{fig02}). The lattice has additional dispersive element  (injection chicane \footnote{The laser heater at FLASH uses the originally proposed layout \cite{lh} with the undulator on the straight path and a chicane in front of it for in-coupling of the laser radiation.} of the laser heater \cite{lh}) which has to be taken into account when looking for working points with a desired global compression. In the current design the  momentum compaction factor $R_{56}^{LH}$ of the laser heater is equal to -2.5 mm. The second order optics gives  the second order momentum compaction factor of the laser heater  $T_{566}^{LH}$  equal to 3.7 mm (see Eq.~(\ref{Eq_si}) for the definition of $R_{56}$ and $T_{566}$). We will use these values in order to correct the momentum compaction factors of the first compression stage in the simple analytical model described below.

The electron bunches, each with the charge of 250 pC, are  produced by shaped laser pulses in the RF gun. The parameters used in the gun simulations for two bunches are listed in Table~\ref{TableInj}. The RF phase is given relative to the phase of maximal mean momentum.

\begin{table}[htbp]
	\centering
	\caption{The injector parameters.}
	\label{TableInj}
	\begin{tabular}{lcccll}
		\hline\hline
		{\bf subsection}& 	{\bf parameter}&	{\bf A-bunch}& 	{\bf S-bunch}\\
		\hline
		{\bf laser} &rms length, ps& 4 & 4\\
		&apperture diameter, mm& 1.15& 1.15\\	                   
		{\bf RF cavity} &frequency, GHz& 1.3 &  1.3\\
		&maximal field on cathode, MV/m& 45& 45 \\	                   
		& relative phase, degree&- 10 & 10\\      
		{\bf solenoid} &Magnetic field, T& 0.172 & 0.172\\
		\hline\hline
	\end{tabular}
\end{table}

\begin{figure}[tb]
	\includegraphics[width=1.0\textwidth]{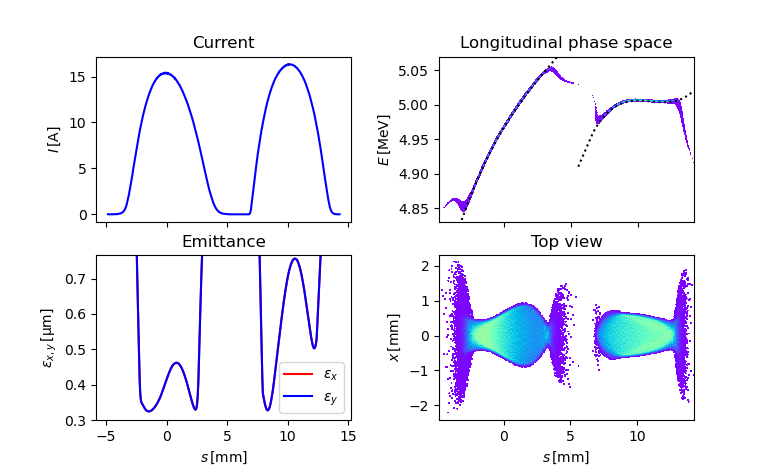}
	\caption{
		The properties of the two bunches at distance $z=2.6$ m from the cathode of RF gun. The dotted line in the longitudinal phase space corresponds to Eq.~(\ref{Eq_LPS}).
	}
	\label{fig04}
\end{figure}

The simulations are done with code ASTRA~\cite{ASTRA}. In our simulations  we use $2\cdot 10^6$ macro-particles per bunch. The slice parameters and the phase space projections at the distance 2.6 meters from the cathode are shown in Fig.~\ref{fig04}.  In order to calculate the slice parameters we have used 5e3 particles per slice. The "emittance" means in this paper the  rms  normalized emittance. 

\begin{figure}[tb]
	\includegraphics[width=1.0\textwidth]{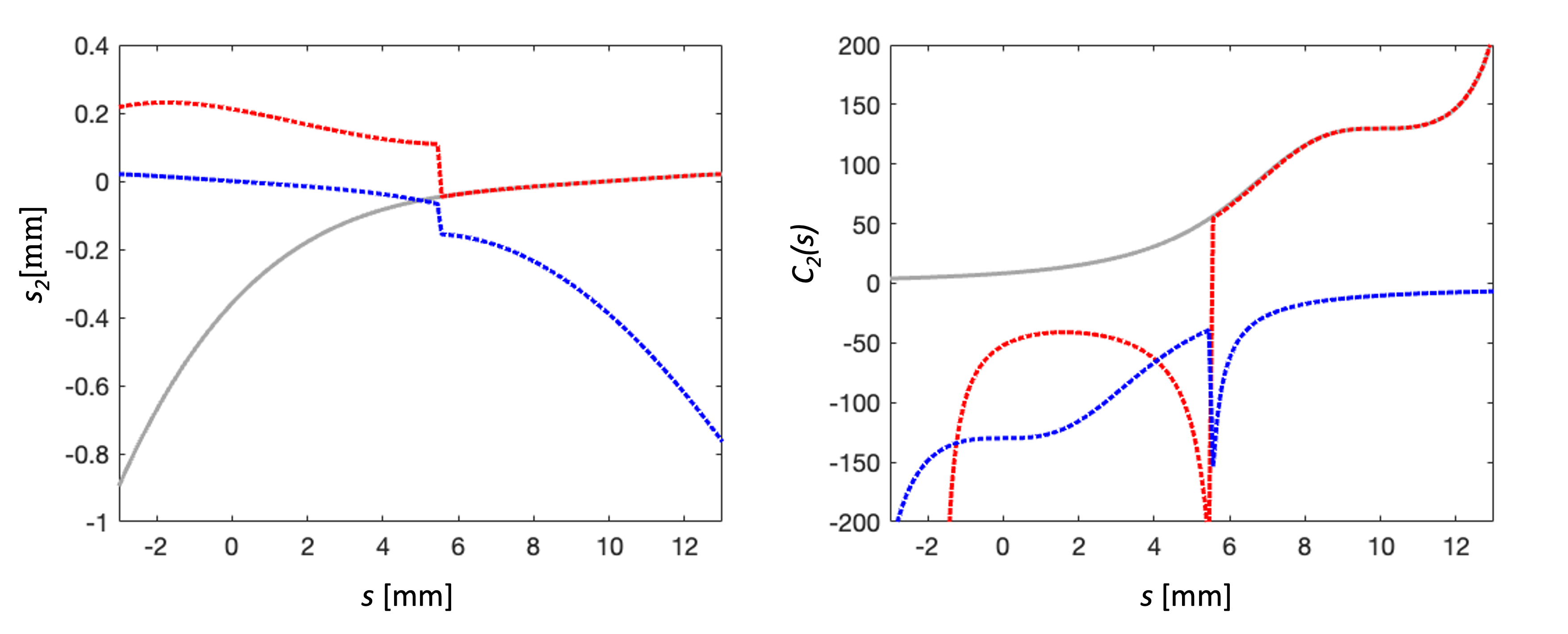}
	\caption{
		The position (on the left) and the compression (on the right) curves of the longitudinal dynamics without collective effects. The gray solid curve describes the position (on the left plot)  and the  compression (on the right plot) for the ideal beam without energy spread. The dashed curves present the position and the compression for the longitudinal phase space approximated by Eq.~(\ref{Eq_LPS}).
	}
	\label{fig05}
\end{figure}

In order to describe the longitudinal beam dynamics inside the bunch during the compression and acceleration let us introduce several coordinate systems. As a starting point we consider the two bunches of electrons after the RF-gun at position $z=2.6$ m. The relative coordinate along the bunches will be noted as $s$.  
It has an origin at the position of the peak current at the A-bunch and increases in the direction of the bunch motion: the head of the bunch has a positive value of $s$, the tail has a negative one. The peak current in the S-bunch is at the position $s=10$ mm.

The longitudinal phase space of A-bunch can be approximated  locally by the third order polynomial
\begin{align}
	E_{0}^A(s)=E_0^{ref,A} (1+ \zeta_{1}^A s+ \zeta_{2}^A s^2+\zeta_{3}^A s^3),
\end{align}
with the coefficients from the first column of Table~\ref{table_L}. The longitudinal phase space of S-bunch can be approximated  locally by the third order polynomial
\begin{align}
	E_{0}^S(s)=E_0^{ref,S} (1+ \zeta_{1}^S (s-0.01 \text{m})+ \zeta_{2}^S (s-0.01 \text{m})^2+\zeta_{3}^S (s-0.01 \text{m})^3),
\end{align}
with the coefficients from the second column of Table~\ref{table_L}. 

Hence after the RF gun the mean slice energy can be approximated by step-wise function		
\begin{align}
		\label{Eq_LPS}
 E_0(s)=
\begin{cases}
	E_0^A(s), & s\leq 0.0055 \text{m}, \\
    E_0^S(s), & s>0.0055  \text{m},\\
\end{cases}
\end{align}
shown by the dotted curve in Fig.~\ref{fig04}.

For the fixed reference energy $E_0^{ref}$ we introduce the relative energy deviation coordinate $\delta_0(s)=(E(s)-E_0^{ref})/E_0^{ref}$. 

\begin{table}[htbp]
	\centering
	\caption{The longitudinal phase space bunch parameters after the RF gun at distance $z$=2.6 m from cathode.}
	\label{table_L}
	\begin{tabular}{lcccll}
		\hline\hline
		{\bf parameter}& 	{\bf A-bunch}& 	{\bf S-bunch}\\
		\hline
		\boldmath $E_0^{ref}$, MeV &4.97&5.01\\
		\boldmath $\zeta_{1}$,  1/m &6.42&-0.14\\
		\boldmath $\zeta_{2}$, 1/m/m&-420&-390\\
		\boldmath $\zeta_{3}$, 1/m/m/m &38590&140748\\
		\hline\hline
	\end{tabular}
\end{table}

Let us consider the transformation of the longitudinal phase space distribution in a multistage bunch compression and accelerating system shown in Fig.~\ref{fig02}. The system has  two bunch compressors $\{BC_1, BC_2\}$ and several   accelerating sections $\{L_1, L_2, L_3\}$. The injector section $L_1$ includes the booster and the third harmonic module. 

In order to describe the longitudinal beam dynamics we introduce several additional reference points to already described one (after the RF gun). The longitudinal coordinate after bunch compressor number $i$  will be denoted as $s_i$ , the energy coordinate at the position immediately  after the bunch compressor will be denoted as $\delta_i$. The reference particle is always at the position $s_i^{ref}$. The coordinate  $s$ (position in the bunch after the RF gun) will be used as an independent coordinate. All other functions depend on it. For example, the function $s_i(s)$ means that the particle with the initial position $s$ (in the bunch after the RF gun) has the position $s_i$  after bunch compressor $BC_{i}$ . In the following we omit the dependence on coordinate $s$  in the notation.

For relativistic electrons, interacting with sinusoidally time varying field, the energy gain of the electron is proportional to the cosine of the phase angle between its position and the position of maximum energy gain. Hence, the energy changes in the accelerating sections can be approximated as
\begin{align}
		\label{Eq_delta}
	\Delta E_{11}&=eV_{11}\cos(ks+\varphi_{11}),\qquad \Delta E_{13}=eV_{13}\cos(3ks+\varphi_{13}),\nonumber\\
	\Delta E_{2}&=eV_{2}\cos(ks+\varphi_{2}), 
\end{align}				
where  $e$ is the electron charge,  $\varphi_{11},V_{11}$ are a phase and  an on-crest voltage of the booster,  $\varphi_{13},V_{13}$ are a phase and  an on-crest voltage of the third harmonic module, $\varphi_{2},V_{2}$ are a phase and  an on-crest voltage of accelerating section  $L_{2}$, and $k$  is a wave number. 

The relative energy deviations in the reference points after the bunch compressors read
\begin{align}
	\delta_1&=\frac{(1+\delta_0)E_0^{ref}+\Delta E_{11}+\Delta E_{13}}{E_1^{ref}}-1,\nonumber\\
	\delta_2&=\frac{(1+\delta_{1})E_{1}^{ref}+\Delta E_{2}}{E_2^{ref}}-1,
\end{align}	
The transformation of the longitudinal coordinate in compressor number $i$  can be approximated by the expression 
\begin{align}
	\label{Eq_si}
	s_i=s_{i-1}-(R_{56i}\delta_i+T_{566i}\delta_i^2+U_{5666i}\delta_i^3),\quad i=1,2.
\end{align}	

\noindent where $R_{56i}$, $T_{566i}$, and $U_{5666i}$ are the first, the second and the third order momentum compaction factors of the corresponding chicanes.
Equations (\ref{Eq_LPS})-(\ref{Eq_si}) present a simple non-linear model of multistage bunch compression system. 

For the fixed values of RF parameters and momentum compaction factors we define the compression functions in each bunch compressor: 
\begin{align*}
	C_i(s)=\frac{1}{Z_i(s)}, \quad Z_i(s)=\frac{\partial s_i(s)}{\partial s},\quad i=1,2.
\end{align*}
The global compression function  $C_2(s)$ presents the compression after compressor $BC_2$ which is obtained for the particles in neighborhood of position $s$ (the position in the bunch after the RF gun). For example, if we would like to increase the peak current by factor 50 at the position of the reference particle, then $C_2(s^{ref})=50$. In other words function $C_2(s)$  describes the increase of the current in the slice with initial position $s$.

The energies at the beam compressors are fixed by design studies and are listed in Table~\ref{Table_LD}. The global compression $C_2$ comes from the requirement on the peak current value of 2 kA.

\begin{table}[htbp]
	\centering
	\caption{The longitudinal dynamics parameters.}
	\label{Table_LD}
	\begin{tabular}{lcccccccc}
		\hline\hline
		\boldmath $ E_1^{ref}$ &	\boldmath $(R_{56})_1$, &
		\boldmath $C_1$  &	\boldmath $E_2^{ref}$ &
		\boldmath $(R_{56})_2$&	\boldmath $C_2$ &
		\boldmath$C_2^{\prime}$  &	\boldmath	$C_2^{\prime\prime}$\\	
		{MeV} &	 {mm} & &	 {MeV} &
		{mm} &	&
		{1/m} & {1/m/m}\\
		\hline
	    130 &	-122 &	3&550 &-104&-130 &	0 & 0\\
		\hline\hline
	\end{tabular}
\end{table}

\begin{table}[htbp]
	\centering
	\caption{The RF parameters.}
	\label{Table_RF}
	\begin{tabular}{lcccccccc}
		\hline\hline
		&\boldmath $V_{11}$ &	\boldmath $\varphi_{11}$  &
		\boldmath $V_{13}$  &	\boldmath $\varphi_{13}$  &
		\boldmath $V_{2}$  &	\boldmath $\varphi_{2}$  \\	
		& {MV} &	 { deg} &
		{ MV} &	 {deg} &
		{MV} &	 {deg} \\
		\hline
		{\bf analytical}	&146.4 &	5.66 &	21.0&168.53&436.8&15.95 \\
		{\bf with self-fields }&144.8&	-2.35 &22.5&150.42 &	438.8&16.23 \\
		\hline\hline
	\end{tabular}
\end{table}

The first and the second derivatives of the global compression are special parameters which allow to tune the flatness and the symmetry of the current profile. We put them  to zeros.  It means that we would like to have a flat current profile  at the vicinity of the reference position. 
 
 The introduced  one dimensional model of the longitudinal beam dynamics neglects the collective effects and the velocity bunching. In order to find the RF parameters of the accelerating modules we use the analytical solution published in ~\cite{Zag11}. 
 
 Let us assume first  that the lasing bunch is the first bunch at the position $s^{ref}=10$ mm. If we assume additionally that the both bunches have no energy chirp and the longitudinal phase space can be approximated by a constant line $E_0(s)=E_0^{ref,S}=5.01$ MeV then the both bunches can be smoothly compressed according to the compression curve shown by the solid gray line in Fig.~\ref{fig05}. The first bunch at the reference position $s^{ref}=10$ mm is compressed by factor $C_2(s_{ref})=130$. The second bunch at the position $s=0$ mm is compressed by the low factor $C_2(0)=8.4$. The bunches arrive to the end of the linac at the same order as they have been emitted from the gun. Unfortunately the second bunch has not flat longitudinal phase space and for the "real" longitudinal phase space approximated by Eq.~(\ref{Eq_LPS}) the compression curve is different. It is shown by dotted red curve in Fig.~\ref{fig05}. It can be seen that the second bunch is overcompressed: $C_2(0)=-52$. In the particle tracking simulations the second bunch arrives as the first to the end of the linac. Hence at the scenario of the flat compression ($C_2(s^{ref})^{\prime}=0, C_2(s^{ref})^{\prime\prime}=0$) the second bunch cannot be used as a seed.
 
 In order to use the first bunch at the position $s^{ref}=10$ mm as a seed we should to work in the overcompression scenario which changes the order of the bunches after $BC_2$. If we require that the second  bunch at the position $s^{ref}=0$ mm is overcompressed with the compression factor $C_2(0)=-130$ then the compression curve has the form shown by the dotted blue line in Fig.~\ref{fig05}. The first bunch is overcompressed by factor $C_2(0.01 \text{m})=-10.3$ and the bunches arrive in the reversed order to the end of the linac. The RF parameters for the last curve are listed in the first raw of Table~\ref{Table_RF}.
 
\begin{figure}[tb]
	\includegraphics[width=1.0\textwidth]{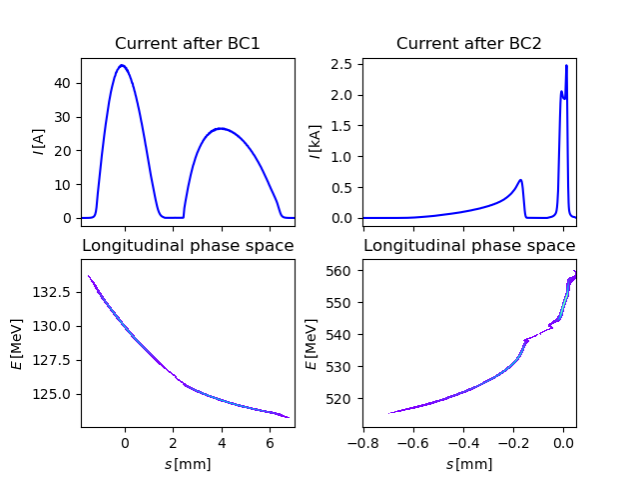}
	\caption{
		The current and the longitudinal phase space after  bunch compressor $BC_1$ (on the left) and after bunch compressor $BC_2$ (on the right).
	}
	\label{fig06}
\end{figure}

The above results are confirmed with the particle tracking in Ocelot with the collective effects included. The RF parameters for this case are listed in the second raw of Table~\ref{Table_RF}.

Fig.~\ref{fig06} presents the current profiles and the longitudinal profiles of the two bunches after the bunch compressors. It can be seen that after the last compressor $BC_2$ the bunches change their order and  both of them go through the overcompresion.

The physical models and the numerical algorithms of code Ocelot~\cite{OCELOT} are described shortly in Appendix B of paper~\cite{Zag19}. The numerical modeling of the accelerator beam dynamics presented in this paper includes the wake functions of the accelerating modules in the form described in ~\cite{Zag20}.  We have tested with the direct numerical solution of Maxwell equations~\cite{ECHO}, that this form of the wake functions describes accurately the wakefields in the second bunch as well.

\begin{figure}[tb]
	\includegraphics[width=1.0\textwidth]{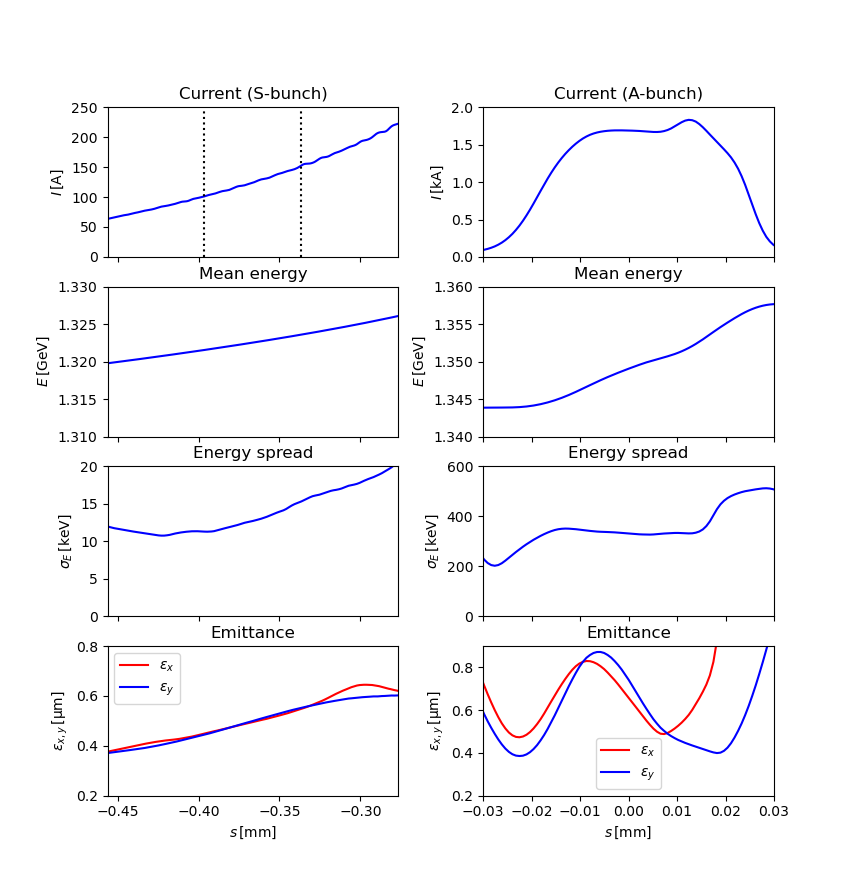}
	\caption{
		The current, the mean slice energy, the slice energy spread and the slice emittance of the seed bunch before the modulator (on the left) and of the lasing bunch before the amplifier (on the right). The dotted lines on the left current plot outline the simulation window used in the FEL modeling.
	}
	\label{fig07}
\end{figure}

Fig.~\ref{fig07} shows the slice parameters of the S-bunch before the modulator (on the left) and of the A-bunch  before the amplifier (on the right).  The slice parameters of each bunch are shown for  the simulation window of $60~\mu$m used in the FEL simulations described below. The peak current of the A-bunch is reduced by approximately 15 \% in the delay chicane before the amplifier due to the strong energy chirp in  the A-bunch.

\begin{figure}[tb]
        \includegraphics[width=1.0\textwidth]{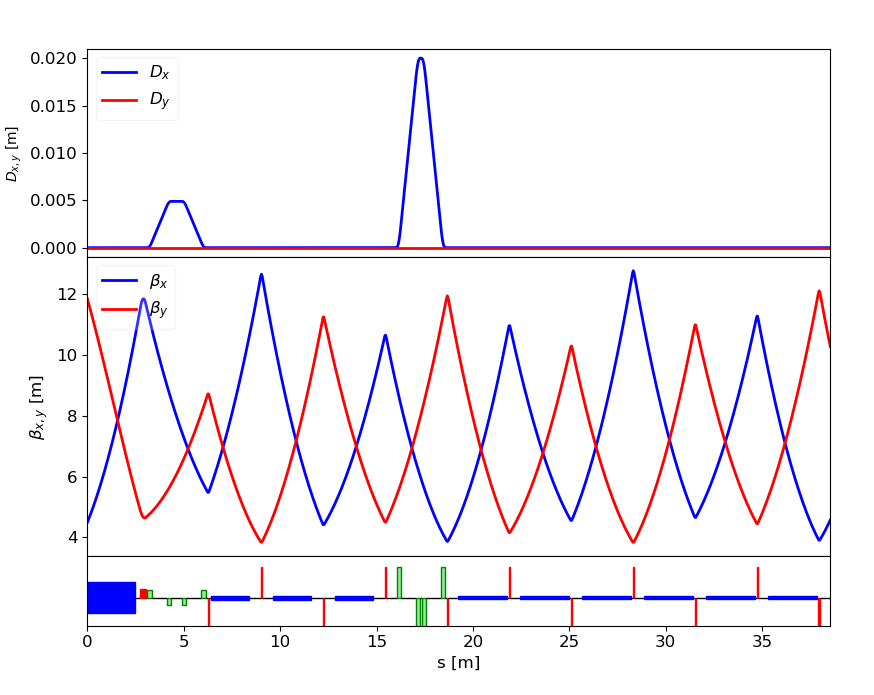}
        \caption{
                The design optics of the seeding setup. The bottom plot presents an outline of different elements: quadrupoles (in red), dipoles (in green), undulators (in blue). The dotted vertical line shows the rematching position for the lasing bunch.
        }
        \label{fig08}
\end{figure}

\section{FEL simulations}

The FEL simulations are carried out with the code Genesis-1.3-Version4~\cite{Reiche} in the optics shown in Fig.~\ref{fig08}. The S-bunch is matched to the entrance of the setup. The same transformation was applied for both bunches. It resulted  in considerable optics mismatch for A-bunch. The mismatched A-bunch was tracked in Ocelot to the position before the amplifier shown by the dotted line in   Fig.~\ref{fig08} and matched to the optics. The Genesis FEL simulations are done with S-bunch up to this position and with A-bunch after it. Hence we assume that it is possible to develop a matching section which will match the A-bunch to the optics of the amplifier section, the details we leave for future studies. The simulations are done with real number of particles, and shot noise is properly included.

\begin{table}[htbp]
	\centering
	\caption{The parameters of FEL simulations.}
	\label{Table_FEL}
	\begin{tabular}{llcc}
		\hline\hline
			{\bf subsystem }& {\bf parameter}& 	{\bf  Value}\\
		\hline	
		Laser& Laser pulse duration (FWHM) , fs &33\\
		&Laser wavelength, nm&300\\
		&Laser power, MW&100\\
		Modulator&Undulator period, cm &8.26\\
		&Number of periods&30\\ 
		&Undulator parameter $K_{rms}$&6.9117\\
		Chicane&  R56, $\mu$m&52\\
		Radiator&Number of modules&3\\
		&Undulator period, cm&3.15\\
		&Number of periods in one module&63\\
		&Undulator parameter $K_{rms}$&0.8400\\
		Delay chicane&  R56, $\mu$m&740\\
		Amplifier&Number of modules&6\\
		&Undulator period, cm&3.5\\
		&Number of periods in one module&72\\
		&Undulator parameter $K_{rms}$&0.7725\\
		\hline\hline
	\end{tabular}
\end{table}

The FEL setup is close to the design of the new undulator line of FLASH1. The only simplification we did in simulations is the removal of a small chicane in the amplifier part (it will be installed later but will not be used in our scheme). The delay chicane between CHG undulator and the amplifier will also not be installed initially. Thus, the two-bunch seeding will be enabled only after installation of that chicane.   

The main parameters used in FEL simulations are presented in Table~\ref{Table_FEL}. The CHG undulator is planar and consists of three segments followed by the delay chicane. The amplifier is place behind the chicane and consists of six segments with variable polarization. We simulate two case: linear and circular polarization of the amplifier undulator. In the latter case only a half of the radiation power from CHG undulator is coupled to the amplifier (linearly polarized beam can be decomposed into left and right circularly polarized beams), so that we introduce the reduction by a factor of two in our simulations.

\begin{figure}[tb]
		
	\includegraphics[width=1.0\textwidth]{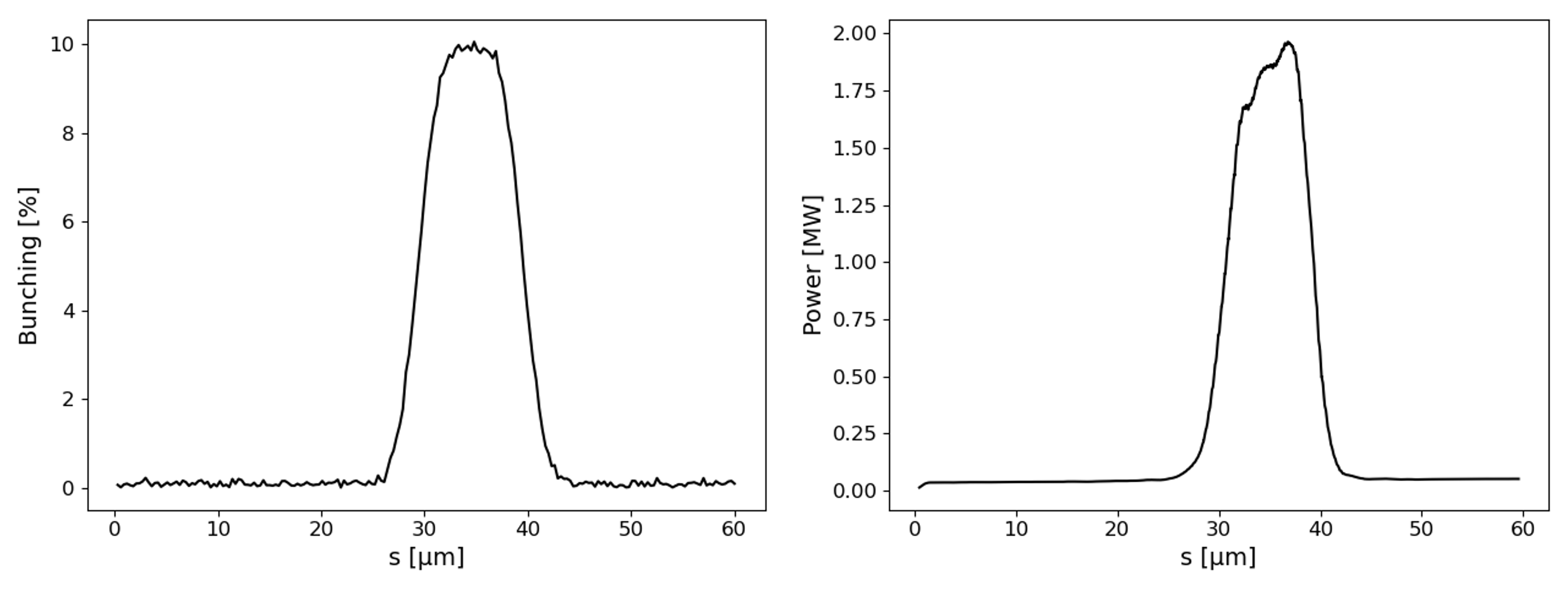}
	
	\caption{
		The left plot shows the bunching factor at the seeding bunch after the undulator. The right plot presents the FEL power profile after radiator.
	}
	\label{fig09}
\end{figure}

The laser pulse is parked on the part of the S-bunch that can be seen on in Fig.~\ref{fig07}. For a given power of the laser pulses (100 MW) we optimize the $R_{56}$ of the chicane. The bunching factor after the modulator-chicane system is shown in Fig.~\ref{fig09}. One can notice a relatively high value of the bunching factor (nearly 10\% at the 75th harmonic of the laser despite its moderate peak power). This is possible due to the small uncorrelated energy spread in the corresponding part of S-bunch, about 10 keV. The radiation pulse at the wavelength of 4 nm, produced in CHG undulator, is also shown in Fig.~\ref{fig09}. The bunching and the power are plotted versus longitudinal coordinate as they are extracted from Genesis simulations, one can easily convert it to time coordinate. Note that during the simulations we optimize the K-value of the CHG undulator.

\begin{figure}[tb]
	
	\includegraphics[width=1.0\textwidth]{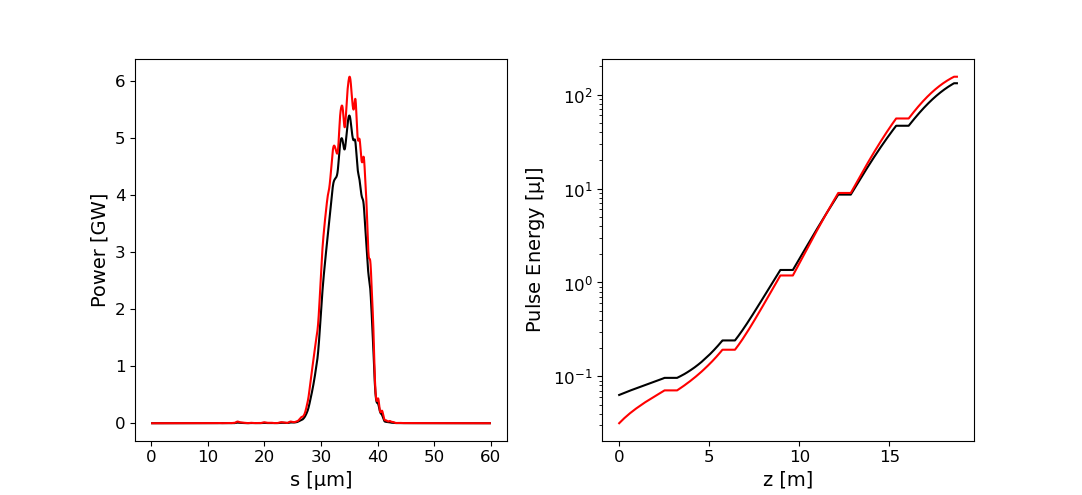}
	
	\caption{
		The left plot shows the output FEL power profiles of planar (black curve) and helical (red curve) undulators. The right plot presents the gain curves of planar (black curve) and helical (red curve) undulators.
	}
	\label{fig10}
\end{figure}

After the CHG undulator the soft X-ray radiation pulse is propagated in a drift space (diffraction is properly included in the simulations) and then it is parked on the part of the A-bunch shown in Fig.~\ref{fig09}. The delay in the chicane is about 0.37 mm (slightly larger than 1 ps). The corresponding $R_{56} = 0.74$ mm is sufficient for a complete smearing of the modulations in S-bunch at 4 nm. Moreover, its current is too low to have any significant FEL gain in the amplifier. The radiation pulses produced by the A-bunch after six segments of the amplifier undulator for linear and circular polarization cases are shown in Fig.~\ref{fig10} along with the corresponding gain curves. Pulses energies at the undulator end are about 150 $\mu J$, pulse durations about 25 fs. Despite in the circularly polarized case the effective input power is lower, the FEL gain and the output power are somewhat higher due to a better coupling between the electron motion and the electromagnetic field \cite{book}. 
Note that we optimized K-value of the amplifier undulator to maximize the FEL power. If the undulator was longer, a higher radiation power could be produced with some post-saturation taper. However, even in the simulated case with six undulator segments we can observe multi-gigawatt peak power thanks to the high peak current of the A-bunch.

\begin{figure}[tb]
	
	\includegraphics[width=0.6\textwidth]{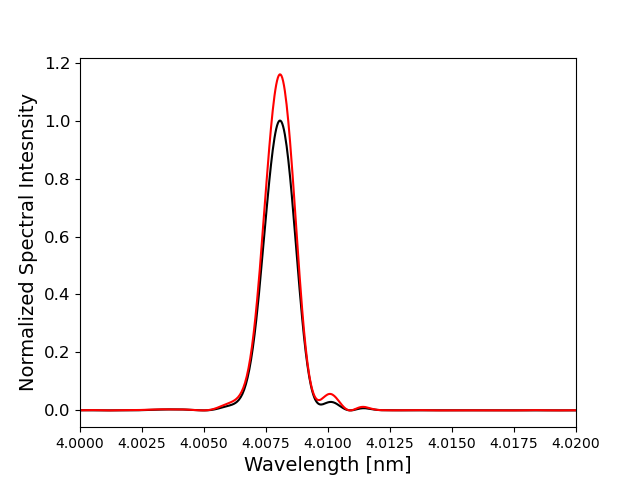}
	
	\caption{
		Output FEL spectra of planar (black curve) and helical (red curve) undulators.
	}
	\label{fig111}
\end{figure}

Output radiation spectra for the cases of linearly and circularly polarized radiation are presented in Fig.~\ref{fig111}. Despite the slight distortions, one can notice a relatively high quality of spectra at the high harmonic number considered in this paper. The relative spectrum width is about $3.4 \times 10^{-4}$ (FWHM), and the time-bandwidth product exceeds Fourier limit for a Gaussian pulse by only $40\%$.

\section{Conclusion}

We proposed two-bunch seeding concept and demonstrated its validity in start-to-end simulations. We showed that nearly Fourier-limited multi-gigawatt pulses can be generated at the wavelength of 4 nm in HGHG configuration with the compact undulator design of FLASH. Some aspects (EEHG case, specific matching optics, sensitivity studies, lasing at even shorter wavelengths, reduction of the required laser power at moderate harmonic numbers) will be studied in future works but already now we can conclude that the concept promises enormous improvements with respect to the traditional seeding scenarios.

\section{Acknowledgments}

The authors would like to thank M. Vogt, J. Zemella and S. Tomin for providing the optical layout shown in Fig~\ref{fig03},
G. Paraskaki for the help with the Genesis 1.3 simulations, E. Ferrari for careful reading of the manuscript and useful comments. We are greatful to W. Leemans, S. Choroba and L. Schaper for their interest in this work.

\end{document}